\documentstyle[12pt]{article}
\begin{document}
\begin{titlepage}
\begin{flushright}
IPNO/TH 94-11 \\
SISSA/27/94/EP \\
UTF/324
\end{flushright}
\hspace{9cm}
\vspace{1cm}
\begin{centering}

{\huge Haldane's Fractional Statistics and the \\
\vspace{0.5cm}
Riemann-Roch Theorem}\\
\vspace{1cm}
{\large Dingping Li$^{a,c}$
and St\'ephane Ouvry$^{b}$}\\
{$^a$Dipartimento di Fisica, Universit\'a di Trento, 38050 Povo, Trento}\\
{$^b$Division de Physique
Th\'eorique\footnote{Unit\'e de Recherche des Universit\'es Paris 11
et Paris 6 associ\'ee au CNRS.},
IPN, 91406 Orsay Cedex and LPTPE,
Universit\'e Paris 6}\\
{$^c$International School for Advanced Studies, SISSA, I-34014
Trieste, Italy}\\
\end{centering}
\vspace{.25cm}
\begin{abstract}

The new definition of fractional statistics given by Haldane
can be understood in some special
cases in terms of the Riemann-Roch theorem.

\end{abstract}
\vspace{1.25cm}
\begin{flushleft}
March, 1994 \\
Submitted to Nucl. Phys. B(FS)
\end{flushleft}

{\footnotesize Electronic address: $^a$LIDP@TSMI19.SISSA.IT,
$^b$OUVRY@IPNVAX.IN2P3.FR}

\end{titlepage}
\def\carre{\vbox{\hrule\hbox{\vrule\kern 3pt
\vbox{\kern 3pt\kern 3pt}\kern 3pt\vrule}\hrule}}


\section{Introduction}

Statistics plays an important role in
the physics of quantum many-body systems.
Bosonic and fermionic statistics have been known
to us for  quite a long time and were believed to be unique.
However,  a few years ago, it was found that,
in two-dimensional systems, fractional statistics,
neither bosonic or fermionic \cite{lei},  can exist.
The particles obeying fractional statistics have been called anyons.

It is now established
that quasiparticles  in the fractional quantum Hall effect
(FQHE)  obey fractional statistics.
Indeed, in ref.\ \cite{halperin},
Halperin conjectured that the fractional quantum
Hall quasiparticles are anyons; he also
suggested
that condensation of such quasiparticles in the Laughlin states
(at the Landau filling $\nu =1/m$)
gives rise  to  hierarchical states ($\nu \not= 1/m$)
(see also ref.\ \cite{haldane1}). The theory of
hierarchical states developed by Halperin and Haldane
is called the standard hierarchical theory
(for a review on the quantum Hall effect,
see Refs.\ \cite{pg,macdonald}).
The Halperin conjecture about anyonic
quasiparticles was soon proved by Arovas, Schrieffer and Wilczek
\cite{arovas}.
Other possible applications of anyon physics, as
anyonic high temperature superconductivity,
can be found in ref.\ \cite{wilczek}.

Recently,  Haldane proposed \cite{haldane}
a new definition of fractional statistics
(NDFS), based on Hilbert space counting argument.
The NDFS has arose a lot of interest
and has been used for  the study of
the FQHE hierarchical states \cite{he,yang,joh}.
It can be viewed as a generalization of the Pauli
exclusion principle in the case of systems with a finite Hilbert
space or subspace. Precisely, the generalization of
the Pauli exclusion principle to fractional
statistics
has been investigated in ref.\ \cite{ouvry}
in the case of anyons in a strong magnetic field, confined to the
infinite Hilbert space of the lowest
Landau level (LLL).
In this situation, it was found that at most $\pi/|\theta|$
anyons can occupy a  given  quantum LLL state
($\theta=0$, Bose statistics, $|\theta | =\pi$, Fermi
statistics, anyonic statistics $|\theta |< \pi$;
notations and conventions will be introduced later),
and that at the critical filling, the magnetic field is
entirely screened by the flux tubes carried by the anyons.
Very recently,  Wu has shown \cite{wu}
how Haldane's Hilbert space counting arguments in a mean field
approach can lead to similar conclusions.
However, Haldane's NDFS can  extend $|\theta  | $ to all
possible values, as we shall see later.

The NDFS can also be used to calculate the size
of the full Hilbert space of many-particle states.
In ref.\ \cite{he,joh}, the emphasis was put on the Hilbert space
of the low-energy sector of fractional quantum Hall states
in the presence of quasielectron (QE) or quasihole (QH) excitations.
On the other hand, the energy spectra of a few electrons can
be calculated numerically. A low-energy sector was found which
is well separated from the groundstate and corresponds
to Hall states with QE or QH excitations. The dimension
of the Hilbert space of such states  predicted in ref.\ \cite{haldane}
is indeed in agreement with the number of states in the low-energy sector.
Furthermore, the low-energy sector has been investigated
in Jain's theory \cite{jain}
of hierarchical states, the so called
composite fermion theory of the FQHE. The number  of
low-energy states can also be obtained in
this framework \cite{dev},
and was found to be identical to
the one predicted in the standard
hierarchical theory using the NDFS concept.
This might give a further evidence of the equivalence
of the two hierarchical theories.

Note finally that the NDFS can in principle help to define statistics in
any dimension and thus suggest a possible
generalization of the notion of fractional statistics to
space of dimension other than two.

However, it is clear that  a complete and convincing picture of
NDFS is still lacking. In view of the several points mentioned above,
it is important  to improve our understanding about it.
We will show that in one
case, conventional anyons
interacting with a strong\footnote{ by strong magnetic field we mean
that one concentrates on the groundstate, meaning that thermal
excitations are negligible compared with the cyclotron gap.} magnetic field,
the new definition is equivalent to the old one.
The Riemann-Roch theorem is used to
demonstrate this equivalence. Several examples (including the FQHE)
will be worked out in detail to illustrate our claim.

\section{Anyons in a strong magnetic field}

We start by a simple example based on ref.\ \cite{li}.
Consider on the sphere $N$ anyons with
hard core boundary conditions interacting with a magnetic field
and use  projective coordinates on the plane, to get
the Landau Hamiltonian\footnote{
The Landau problem for anyons on the plane has been
originally discussed in ref.\ \cite{many}}
 (see also ref.\ \cite{comtet})
\begin{equation}
H=\sum_{i=1}^N H_i
={2\over M}\sum_{i=1}^N(1+z_i\bar z_i)^2
(P_{z_i}-A_{z_i})(P_{\bar z_i}-A_{\bar z_i})  .
\end{equation}
with
\begin{equation}
A_{z_i}=
i {\theta \over 2\pi} \sum_{j\not=i}[{1\over z_i-z_j}]
-i {\phi \over 2}
{\bar z_i\over 1+z_i\bar z_i} .
\label{potential}
\end{equation}
and
\begin{equation}
P_{z_i}=-i \partial_{z_i} \, \, ,P_{\bar z_i}=-i
\partial_{\bar z_i} \, \, .
\end{equation}
($A_{\bar z_i}$ is the complex conjugate of
$A_{z_i}$).
The Hamiltonian with Laplace-Beltrami ordering is
\begin{equation}
H_{L-B}={1\over M}\sum_{i=1}^N(1+z_i\bar z_i)^2
[(P_{z_i}-A_{z_i})(P_{\bar z_i}-A_{\bar z_i})
+ (P_{\bar z_i}-A_{\bar z_i})(P_{z_i}-A_{z_i})]
\end{equation}
with
\begin{equation}
H_{L-B}-H=
\sum_i (1+z_i\bar z_i)^2
({\phi \over (1+z_i\bar z_i)^2} -{\theta} \sum_{i\not= j}
\delta^2 (z_i-z_j)).
\end{equation}
So, to the exception of a constant term and $\delta$ functions which can be
omitted because of hard core boundary conditions
which exclude the diagonal
of the configuration space (i.e. wavefunctions have
to vanish when any $2$ anyons coincide),
$H$ and $H_{L-B}$
are equivalent.

In (2), the first term of $A_{ z_i}$ encodes the fractional statistics
interaction with statistical parameter $\theta /\pi$.
The many-body wavefunction has to be symmetric
under coordinate exchange since by convention the particles obey
bosonic statistics.
If a singular gauge transformation is performed to eliminate this
term,  the many-body wavefunction explicitly obeys fractional statistics
:
the statistical phase factor
is $e^{i\theta}$, thus  the statistical phase is periodic
with period $2\pi$.
It will become soon  clear that  if $\theta$ is shifted by $2\pi$,
many physical quantities will be affected.

The second term in (2) describes the anyons coupling with
the magnetic field. $\phi$ is  related to  $q\Phi$, where
$q$ is the charge of the particle
and $\Phi$ is the magnetic flux
out of the surface.
In the case of FQHE quasiparticles,
$\phi -q\Phi$  is a finite constant
related to  the spin of the particle
(see ref.\ \cite{lispin} and  the following sections).

Now let us fix the coordinates of $z_i, \quad i\not= 1$.
The Hamiltonian of particle $1$  is simply $H_1$.
The Dirac quantization condition for $H_1$ is
\begin{equation}
\phi - (N-1){\theta \over \pi}=n
\label{index}
\end{equation}
with $n$ an integer. In what follows, all parameters, in particular
$\phi$ and $n$,  are assumed to be
positive for convenience.
Other situations,  for example the  QE case,
 will be discussed afterwards.
The Dirac quantization condition can be also presented as follows :
the flux felt by a given particle should be an integer
expressed in magnetic quantum flux unit.

Since $H_1$ is  positive definite, if  one finds $\psi (z_1)$  such
that $H_1\psi (z_1)=0$, it is the groundstate.
At low temperature and in a strong magnetic field
($\phi /M$ is  big compared with the thermal energy), the
system will be confined
to the groundstate, i.e. the LLL, defined by
\begin{equation}
(P_{\bar z_1}-A_{\bar z_1})\psi (z_1)=0.
\end{equation}
Solutions  are found to be
\begin{equation}
\psi (z_1)=(z_1)^k \prod _{i\not= 1}[{(z_1-z_i)
(\bar z_1-\bar z_i)\over (1+z_1 \bar z_1)}
]^{\theta \over 2\pi}
(1+z_1 \bar z_1)^{ -n /2 },
\end{equation}
with $k=0, 1, \cdots , n$.
The dimension $d$ of this  Hilbert subspace (groundstate) is thus equal
to $n+1$.

Before applying Haldane's NDFS to this particular finite dimensional
Hilbert space, let us first recall its general definition.
Consider a $N$-body system, and fix $N-1$ particles among them.
Analyze the Hamiltonian of the remaining particle (for example particle
$1$) assuming  that the dimension $d$ of the Hilbert space for this particle
(or subspace of
the Hilbert space) is finite and independent
of which particle has been chosen.
The NDFS statistical parameter $g$ is by definition
\begin{equation}
\Delta d =-g \Delta N.
\end{equation}
If one considers varying number of particles
in order that a thermodynamic limit can be properly defined,
$\Delta d =-g \Delta N$ is an integer, so is $\Delta N$,
thus $g$ must be rational.

The size of the Hilbert space of the many-body system (bosonic)
at fixed $N$ is then
\begin{equation}
{(d+N-1)! \over (N)!(d-1)!}.
\label{size}
\end{equation}
In the present case,
$d=1+\phi - (N-1){\theta / \pi}$,
one finds $g={\theta / \pi}$ :
the NDFS statistical parameter is thus identical
to the fractional  statistical parameter.
Now one can find the wavefunctions of the many-body state
\begin{eqnarray}
\Psi (z_1,\cdots,z_i,\cdots,z_N)
& = & f(z_i) \prod _{i<j}[{(z_i-z_j)
(\bar z_i-\bar z_j)\over (1+z_i \bar z_i)
(1+z_j \bar z_j)}]^{\theta \over 2\pi} \nonumber \\
& & \times \prod_i (1+z_i \bar z_i)^{-n / 2}  ,
\label{wvfu}
\end{eqnarray}
where $f(z_i)$ is a
symmetric analytic function of $z_i$, with monomial in
$z_i$ of power smaller
or equal to $n$. Symmetric polynomials $\sigma_i$ are generated by
the function
\begin{equation}
P(z_i)=\prod_i (z-z_i)=\sum_{i=0}^N (-1)^i \sigma_i z^{N-i},
\end{equation}
Thus
\begin{equation}
f(z_i)=\prod_{i} {\sigma_i}^{s_i}
\end{equation}
with $\sum_i s_i \leq n$.
Due to this restriction,
the number of $f(z_i)$'s is finite
(so is the dimension of the groundstate)
and is equal to
\begin{equation}
{(N+n)! \over N!n!},
\end{equation}
in agreement with Haldane's equation\ (\ref{size})
(in this case, $d=n+1$).
It can be rewritten as in the mean field approach of ref.\ \cite{wu}
\begin{equation}
{(G+N-1-(N-1)g)!  \over N!(G-1-(N-1)g)!},
\end{equation}
with $G=\phi +1$ ($g=\theta / \pi$). $G$ is the
the number of degenerate Landau states a particle can occupy
($d=G-(N-1){\theta \over \pi}$, thus $G=d$ when $N=1$ as it should).
But $G$ can be non-integer (however $
G-(N-1)g=d$ must be an integer).

If one increases the number of particles, $d$ decreases
to $1$. $d=1$ is critical since below that point one has not enough room
to
put all the  particles in the LLL.
In this critical situation, $\phi -(N-1)\theta/ \pi=0$, the magnetic
field is entirely screened by the flux tubes carried by the anyons.
The critical filling is (in the thermodynamic limit
$N\to\infty$)
\cite{ouvry,wu}
\begin{equation}
\nu ={N \over \phi}={1 \over g}.
\end{equation}
The non degenerate groundstate wavefunction at the critical filling  is simply
\begin{equation}
\Psi_{crit} (z_1,\cdots,z_i,\cdots,z_N)=\prod _{i<j}[{(z_i-z_j)
(\bar z_i-\bar z_j)\over (1+z_i \bar z_i)
(1+z_j \bar z_j)}]^{\theta \over 2\pi}.
\label{wvfucr}
\end{equation}

One can generalize the above results to the case
of multispecies anyons (labelled by
an index $l$)
with mutual statistics \cite{gailuron}.
Now the Hamiltonian is
\begin{equation}
H=\sum  H_{i,l}
=\sum {2\over M_l}(1+z_{i,l}\bar z_{i,l})^2
(P_{z_{i,l}}-A_{z_{i,l}})(P_{\bar z_{i,l}}-A_{\bar z_{i,l}}) ,
\end{equation}
where $z_{i,l}$ is the coordinate of the $i^{th}$ anyon of
species $l$.
The gauge field $A_{z_{i,l}}$ is
\begin{equation}
A_{z_{i,l}}=
i \sum_{j\not=i} {\theta_{l,l} \over 2\pi} [{1\over z_{i,l}-z_{j,l}}]
+i \sum_{j, k\not=l} {\theta_{l,k} \over 2\pi} [{1\over z_{i,l}-z_{j,k}}]
-i {\phi_l \over 2} {\bar z_{i,l}\over 1+z_{i,l}\bar z_{i,l}} .
\label{potential1}
\end{equation}
The Dirac quantization condition for species $l$ reads
\begin{equation}
\phi_l - (N_l-1){\theta_{l,l} \over \pi}
-\sum_{k\not= l}N_k{\theta_{l,k} \over \pi}
=n_l
\label{index1}
\end{equation}
where $N_k$ is the number of  anyons of
species $l$  and
\begin{equation}
d_l=n_l+1.
\end{equation}
One defines
\begin{equation}
\Delta d_l =-\sum_k g_{l,k} \Delta N_k.
\end{equation}
Then
\begin{equation}
g_{l,k} ={\theta_{l,k} \over \pi}.
\end{equation}
The many-body groundstate wavefunctions are
\begin{equation}
\Psi=f(z_{i,l})
\prod_{i, j, l, k}[{(z_{i,l}-z_{j,k})
(\bar z_{i,l}-\bar z_{j,k})\over (1+z_{i,l} \bar z_{i,l})
(1+z_{j,k} \bar z_{j,k})}]^{\theta_{l,k} \over 2\pi}
 \prod_{i,l} (1+z_{i,l} \bar z_{i,l})^{-n_l / 2}
\label{wvfu1}
\end{equation}
where if $l=k$, then $i<j$.
$f(z_{i,l})$ is  an holomorphic function
generated by
\begin{equation}
f(z_{i,l})=\prod_{i,l} {\sigma_i(l)}^{s_{i,l}}
\end{equation}
where the $\sigma_i(l)$'s are symmetric polynomials in the coordinates
$z_{j,l}$. One has also the restrictions
\begin{equation}
\sum_i s_{i,l} \leq n_l.
\end{equation}
The number of solutions is
\begin{equation}
\prod_l {(N_l+n_l)! \over N_l!n_l!},
\end{equation}
or
\begin{equation}
\prod_l {[G_l+N_l-1-\sum_k g_{l,k}(N_k-\delta_{l,k})]!
\over N_l![G_l-1-\sum_k g_{l,k}(N_k-\delta_{l,k})]!},
\end{equation}
with $G_l=\phi_l+1$.

At the critical filling, the $d_l$'s are all equal to $1$
and the groundstate is
non degenerate.

\section{The Riemann-Roch theorem and
the NDFS}

The results of the preceeding section can be easily understood in terms
of the Riemann-Roch theorem. The Riemann-Roch theorem and other
recent developments in Algebraic Geometry
have been used in ref.\ \cite{iengoli}
to investigate Landau
quantum mechanics on various Riemann surfaces.

Following ref.\ \cite{iengoli},
let us define the metric $ds^2=g_{z\bar z}dzd{\bar z}$
where $z$ is a complex coordinate on a given Riemann surface.
The volume form is
$dv=[ig_{z\bar z} / 2]dz\wedge d{\bar z}
= g_{z\bar z}dx\wedge dy$.
Take  a   constant magnetic field applied  perpendicularly to the Riemann
surface, $F=Bdv=(\partial_zA_{\bar z}-
\partial_{\bar z}A_z)dz\wedge d{\bar z}$ implying
$ \partial_zA_{\bar z}-
\partial_{\bar z}A_z =ig_{z\bar z}B / 2$.
The flux of the magnetic field is $2\pi \Phi = \int F =BV$,
where $V$ is the area of the surface and
$B>0$ has been assumed ($\Phi >0$; $\Phi$ should be an integer
because of the Dirac quantization condition). The Landau Hamiltonian
reads
\begin{eqnarray}
H_{L-B} & = & [1/ 2M \sqrt{g}]
(P_{\mu}-A_{\mu})g^{\mu \nu}\sqrt{g}(P_{\nu}-A_{\nu})
\nonumber \\
& = & [ g^{z\bar z} / M]
[(P_z-A_z)(P_{\bar z}-A_{\bar z})  +
(P_{\bar z}-A_{\bar z})(P_z-A_z)] \nonumber  \\
\label{hamil}
& = & [2g^{z\bar z}/ M]
(P_z-A_z)(P_{\bar z}-A_{\bar z})+ B/ 2M
\end{eqnarray}
where $g^{z\bar z}=[1 / g_{z\bar z}]$ and
$P_{z}=-i\partial_z ,\, P_{\bar z}=-i\partial_{\bar z}$.
The inner product is defined as
$<\psi_1 | \psi_2 >=\int dv {\bar \psi_1 }\times \psi_2$.
$H_{L-B}-B/2M$
is a positive definite hermitian operator
thus $(P_{\bar z}-A_{\bar z})\psi =0$ has for solutions
the groundstate wavefunctions of the Hamiltonian $H_{L-B}$,
i.e.  the LLL.
The existence of such solutions
is guaranteed by the Riemann-Roch theorem.
They belong to the
holomorphic line bundle under the gauge field.
The Riemann-Roch theorem states that
$h^0(L)-h^1(L)=deg(L)-h+1$,
where $h^0(L)$ is the dimension of the holomorphic line bundle
or  the degeneracy of the groundstate of $H_{L-B}$,
$h^1(L)$ is the dimension of the holomorphic line bundle
$K\cdot L^{-1}$ where $K$ is the canonical bundle, and
$h$ is the genus of the surface.
$deg(L)$ is the degree of the line bundle which is equal to
the first Chern number of the gauge field, or the magnetic flux
through the surface, $\Phi$.
If $deg(L) >2h-2$ (case of a strong magnetic field as
in the FQHE),  $h^1(L)=0$ and $h^0(L)=\Phi-h+1$.

Now, consider on the surface $N$ anyons with hard-core
boundary conditions
coupled to a constant magnetic field.
The Hamiltonian becomes
\begin{eqnarray}
H_{L-B} =\sum_i  [ g^{z_i\bar z_i} / M]
[(P_{z_i}-A_{z_i})(P_{\bar z_i}-A_{\bar z_i})
 +   (P_{\bar z_i}-A_{\bar z_i})(P_{z_i}-A_{z_i})].
\end{eqnarray}
The vector potential
$A_{z_i}$ is a sum of two terms $A_{z_i}^1+A_{z_i}^2$.
$A_{z_i}^1$ encodes the fractional statistics interaction between
the anyons and $A_{z_i}^2$ describes the coupling of the
anyons with the constant external magnetic field. The flux
is $n=\phi -(N-1){\theta / \pi}$,
where $\phi$ is the contribution of  $A_{z_i}^2$
and $-(N-1){\theta / \pi}$ is those of
$A_{z_i}^1$.
$n$ must be an integer, which is assumed to be positive,
as well as $\phi$ and $\theta$ which are also positive.

Define
\begin{equation}
H=\sum_i [2 g^{z_i\bar z_i} / M]
[(P_{z_i}-A_{z_i})(P_{\bar z_i}-A_{\bar z_i}).
\end{equation}
$H_{L-B}-H$ contains only $\delta$
interactions and constant terms as for the case of the sphere.
Because of  hard core  boundary conditions
these two Hamiltonians are in fact equivalent.
After fixing the coordinates $z_i$, $i\not= 1$, one analyzes
the Hamiltonian of particle $1$.
If some solutions of the equation
$(P_{\bar z_i}-A_{\bar z_i})\psi (z_1)=0$ exist,
they are the groundstate wavefunctions of  $H_1$.
If $n>2h-2$, and following the previous discussions,
the number of solutions $\psi (z_1)$ is $d=n-h+1$
(in the case of the sphere,  $h=0$, then $d=n+1$ as it should).
One obtains
$g=\theta /\pi$ as in the previous section.
Because  the many-body wavefunctions are bosonic
in the gauge chosen,
the size of the Hilbert space is again given by
$(N+d-1)!/[N!(d-1)!]$.
Also,
note that the Riemann-Roch theorem actually proves
that the groundstate wavefunctions found in the last section are
complete.

The above discussion
applies to other various situations.
In the case of a strong magnetic field and at low
temperature,
the system will be confined in the groundstate
$(P_{\bar z_i}-A_{\bar z_i})\psi (z_1)=0$.
The Riemann-Roch theorem does not depend
on details of the interactions, as for example the exact extension or
profile of the vortex
described by the gauge potential $A_{z_i}^1$
(here we have specialized to
point-like flux tubes),
but only  on the flux $n$ stemming from them.

One has only discussed compact surfaces so far,
but real samples do have boundaries.
Then the Riemann-Roch theorem with boundaries \cite{egh}
(or more generally, the index theorem with boundaries)
should be relevant.
One can show that, at the condition that the boundary conditions
are unchanged when particles are added,
$\Delta d$ is still  equal to $-({\theta /\pi})\Delta N$
and thus $g$ is equal to ${\theta /\pi}$.

\section{the FQHE and Haldane's statistics}

Following the original motivation of Haldane
\cite{haldane}, let us use the NDFS concept for
the FQHE quasiparticles excitations.

We remind that the Hamiltonian of an electron on a sphere coupled to
a magnetic field is
\begin{equation}
H={2\over M_e}{(1+z \bar z)}^2(P_z-A_{z})
(P_{\bar z}-A_{\bar z}) ,
\end{equation}
with
\begin{equation}
eA_{z}=-{i \Phi \over 2}
{\bar z \over 1+z \bar z} ,
\end{equation}
where $\Phi$ is the magnetic flux out of the surface.
Define
\begin{equation}
d_{ij}={z_i-z_j \over
(1+z_i \bar z_i)^{1\over 2}(1+z_j \bar z_j)^{1\over 2}}.
\end{equation}
The Laughlin wavefunctions at filling $\nu =1/m$
\cite{laughlin,haldane1,li,lispin} then read
\begin{equation}
\prod_{i<j}^{N_e}d_{ij}^m
\end{equation}
where $N_e$ is the number of electrons.
The Hamiltonian for the quasiparticles is \cite{lispin}
\begin{eqnarray}
H_{qh} & = & {2\over M_{qh}}\sum_{i=1}^{N_{qh}}
\left(1+z_i\bar z_i\right)^2
(P_{\bar z_i} -A_{qh}(\bar z_i))
(P_{z_i}-A_{qh}(z_i))  , \nonumber \\
H_{qe}& = & {2\over M_{qe}}\sum_{i=1}^{N_{qe}}\left(1+z_i\bar z_i\right)^2
(P_{z_i}-A_{qe}(z_i)) (P_{\bar z_i} -A_{qe}(\bar z_i))
\label{hamqh}
\end{eqnarray}
where $N_{qh}$ ($N_{qe}$) is the number of QH (QE).
Different normal orderings for the QH and QE Hamiltonians
will be justified later. However,
because of the hard core boundary conditions on the
quasiparticles \cite{he}, Hamiltonians with different
normal orderings are equivalent.
One has
\begin{eqnarray}
A_{qh}(z_i)& = &{-i\over 2m }\sum_{j\not=i}
{1\over z_i-z_j}-{i\over 2}({1\over m}-1)
{{\bar z_i}\over 1+z_i\bar z_i}+{i\Phi \over 2m}
{{\bar z_i}\over 1+z_i\bar z_i} ,
\nonumber \\
A_{qe}(z_i)& = &{-i\over 2m }\sum_{j\not=i}
{1\over z_i-z_j}-{i\over 2}({1\over m}+1)
{{\bar z_i}\over 1+z_i\bar z_i}-{i\Phi \over 2m}
{{\bar z_i}\over 1+z_i\bar z_i}
\end{eqnarray}
$A_{qh}(z_i)$ was obtained in ref.\ \cite{lispin}, whereas
$A_{qe}(z_i)$  can be obtained in the same way
by calculating the Berry phase of quasiparticles on the sphere
\cite{arovas}.
$M_{qh}$ and $ M_{qe}$ are just arbitrary scales which
do not represent any physical quantities.
In the FQHE, the quasiparticles obey a vortex dynamics,
rather than the Newtonian dynamics of massive particles
\cite{haldane}. The ``guiding center" coordinates do not commute
and such systems  are described by  the quasiparticles Hamiltonian
projected on the LLL.
The  QH and QE statistical phases  are
both equal to $e^{-i\pi /m}$. The  QH charge   is $1/m$
(assuming that  the charge of an  electron
is $-1$) and thus the  QE charge   is $-1/m$.
The first term in $A_{qh}(z_i)$ or $A_{qh}(z_i)$
is due to fractional statistics.
 The last term in $A_{qh}(z_i)$ or $A_{qh}(z_i)$
describes the coupling of  the quasiparticles
to the magnetic field.
The remaining term in $A_{qh}(z_i)$ or $A_{qh}(z_i)$
is related to the intrinsic spin of  the quasiparticles
\cite{lispin}.
For completeness,
the effective Coulomb interactions between quasiparticles should be included,
and they actually are taken into account
via the hard core boundary conditions on the
quasiparticles (we will comment on this later).

Let us first  consider the one-body QH Hamiltonian.
The flux quantization is
\begin{equation}
-{\Phi \over m}+({1 \over m}-1)+{(N_{qh}-1) \over m}=-N_{e}
\end{equation}
where the relation $m(N_e-1)+N_{qh}=\Phi$ has been used.
Because of the negative flux $-N_{e}$,
the groundstate wavefunctions satisfy
$(P_{ z_i}-A_{z_i})\Psi =0$, and not
$(P_{\bar z_i}-A_{\bar z_i})\Psi =0$.
As a special normal ordering as been used in the QH Hamiltonian,
the QH groundstate energy is $0$
(the same reasoning also applies to QE).
If we would include Coulomb interactions,
the energies of those eigenstates would be different
from each other and form a band.

$d$ is now equal to $N_{e}+1$. The QH wavefunctions
can be constructed following
the discussions of section 2
\begin{eqnarray}
\Psi(z_1,\cdots,z_i,\cdots,z_{N_{qh}})
& = & f(\bar z_i) \prod _{i<j}^{N_{qh}}
[{(z_i-z_j) (\bar z_i-\bar z_j)\over (1+z_i \bar z_i)
(1+z_j \bar z_j)}]^{1/2m} \nonumber \\
& & \times \prod_i^{N_{qh}}
(1+z_i \bar z_i)^{-N_e / 2}  ,
\label{wvfuqh}
\end{eqnarray}
with degeneracy equal to
\begin{equation}
{(d-1+N_{qh})! \over (d-1)!N_{qh}!},
\end{equation}
The NDFS parameter $g$  is equal to $1/m$,
according to the previous discussion in section $2$. The
$f(\bar z_i)$ are again symmetric polynomials. At the critical filling,
$d$ is required to be $1$. But it means
that $N_e=0$, which is not physically  interesting.
When the QH condense and  the QH wavefunction
is a Laughlin wavefunction type,
the FQHE state is  a hierarchical state.
The QH wavefunction  at $d=1$ is
\begin{equation}
 \prod_{i<j}^{N_{qh}}
|d_{ij}|^{1/m}(\bar d_{ij})^p,
\label{laugqh}
\end{equation}
where $p$ is a positive even integer.
As in the case of Laughlin states formed by electrons,
Coulomb interactions between quasiparticles are needed
to form the incompressible state (\ref{laugqh}).  Because the
charge of the quasiparticle is $1/ m$, the Coulomb interaction
between quasiparticles is weaker than
the Coulomb interaction between electrons.
Then one should expect that such states are difficult to obtain.
However, we will show that in the QE case,
 fractional statistics interactions play
a significant role to produce the Laughlin states.
In fact, it will come out that QE  Laughlin states are
easier to form than QH Laughlin states.
This conclusion was used in ref.\ \cite{he}
to explain the QH and QE
states asymmetry.

The  flux quantization condition for the one-body  QE Hamiltonian is
\begin{equation}
{\Phi \over m}+({1 \over m}+1)+{(N_{qe}-1) \over m}=N_{e}
\end{equation}
where the relation $m(N_e-1)-N_{qe}=\Phi$ has been used.
The  one-body QE groundstate
(the coordinates of the  other QE are fixed)
is determined by
\begin{equation}
(P_{\bar z_1}-A_{qe}(\bar z_1)) \psi (z_1)=0,
\end{equation}
The solutions are
\begin{equation}
\psi (z_1)=f(z_1) \prod _{i\not= 1}^{N_{qe}}
[{(z_1-z_i)
(\bar z_1-\bar z_i)\over (1+z_1 \bar z_1)}]^{-1 \over 2 m}
(1+z_1 \bar z_1)^{ -N_e / 2 }.
\end{equation}
with $f(z_1)= z_1^k,  \,  k=0, 1, \cdots , N_e$.
It appears that the number of solutions is $d=N_e+1$,
in agreement with the Riemann-Roch theorem.
However, due to the hard core boundary conditions
imposed on the quasiparticles,  only a subset of the above solutions
can be retained
(note that hard core boundary conditions are automatically satisfied
for QH wavefunctions).
Indeed,
$f(z_1)$ should have zeros at $z_1=z_i, \quad i\not= 1$.
But as the many-body wavefunctions should obey bosonic statistics,
the order of zeroes at $z_1=z_i, \quad i\not= 1$ must be
equal or greater than two.
$f(z_1)$ should thus take the form
$f(z_1)=\prod_{j \not= 1} (z_1-z_j)^2 f^{\prime}(z_1)$,
with $f^{\prime}(z_1)=z^k, k=1, 2, \cdots, N_e-2(N_{qe}-1)$.
We conclude that the number of wavefunctions with
hard core boundary conditions is
$d^{\prime}=N_e-2(N_{qe}-1)+1$.
This result can also be understood by
the Riemann-Roch theorem.
The number of zeros of the wavefunctions
is equal to the flux out of the surface, or the Chern number
(in this case the number of zeros or the Chern number is $N_e$).
Now, zeros are at $z_1=z_i, \quad i\not= 1$ with at least order
two (the total order of zeros at these points is  at least
$2(N_{qe}-1)$)).
The number of  such linear independent functions
is  $N_e-2(N_{qe}-1)+1$, and on higher  genus surfaces,
$N_e-2(N_{qe}-1)-h +1$,  by the Riemann-Roch theorem.

Because of the hard core boundary conditions,
the dimension of the one-body  QE Hilbert space
is $d^{\prime}$.
Since $d^{\prime}={\Phi \over m}+({1 \over m}+1)
-(N_{qe}-1)(2-  {1\over m}) $, the NDFS parameter
is found to be $g=2-  {1\over m}$
(this result was also numerically obtained
in ref.\ \cite{joh}).
Thus  the critical filling at $d^{\prime}=1$ is
$1/(2-1/m)$.

The dimension of the many-body QE Hilbert space is
\begin{equation}
{(d^{\prime}-1+N_{qe})! \over (d^{\prime}-1)!N_{qe}!},
\label{sizeqe}
\end{equation}
and the construction of the QE many-body wavefunctions
proceed as in the case of QH :
\begin{eqnarray}
\Psi (z_1,\cdots,z_i,\cdots,z_{N_{qe}})
&  = &  f^{\prime}(z_i) \prod _{i<j}^{N_{qe}}
[{(z_i-z_j)(\bar z_i-\bar z_j)\over (1+z_i \bar z_i)
(1+z_j \bar z_j)}]^{-1/2m} \nonumber \\ & &
\times [{(z_i-z_j) \over  (1+z_i \bar z_i)^{1/2}
(1+z_j \bar z_j)^{1/2}}]^2 \nonumber \\ & &
\times \prod_i (1+z_i \bar z_i)^{-N_e+2(N_{qe}-1) \over 2}
\label{wvfuqe}
\end{eqnarray}
where the $f^{\prime}(\bar z_i)$'s stand for
symmetric polynomials (their construction
has been discussed in section $2$)
and  where the highest power
of any coordinate is $N_e-2(N_{qe}-1)$.
The number of such polynomials is
indeed equal to (\ref{sizeqe}).
The critical filling occurs at $d^{\prime}=1$, which implies
$N_e=2(N_{qe}-1)$ and corresponds to the hierarchical
states with the electron filling  at $1/(m-1/2)$.

The QE wavefunction at $d^{\prime}=1$ is then
\begin{equation}
\prod_{i<j}^{N_{qe}}
|d_{i,j}|^{-1/m}d_{i,j}^2.
\label{laughqe}
\end{equation}
Contrary to the QH Laughlin state (\ref{laugqh}),
the statistical interaction is largely responsible for
the QE Laughlin state (\ref{laughqe}), at the critical filling.
Coulomb interactions do not need to be very strong
to form such a QE Laughlin state.
Thus the QE Laughlin state of (\ref{laughqe})
should be observed relatively easily.  This explains why
there are more FQHE hierarchical states due to
QE condensation than
to QH condensation.
By no means,   one should conclude that Coulomb
interactions are not necessary
to form QE Laughlin states.
The physical origin of hard core boundary conditions for
quasiparticles can actually be found in
the Coulomb short-range repulsion
between  quasiparticles \cite{he}.

The above discussion also shows that
the statistical phase does not uniquely determine
the NDFS parameter.  The QE and QH statistical phases are
the same, but the NDFS parameters for QH and QE
are different.

In the hierarchical theory of Haldane and Halperin,
the filling of the $n$-level
hierarchical  states is
\begin{equation}
\nu ={1\over \displaystyle p_1-
{\strut 1\over \displaystyle p_2-
{\strut 1\over \displaystyle \cdots -
{\strut 1\over \displaystyle p_n}}}}   ,
\label{shier}
\end{equation}
where $p_1$ is a positive integer, and the $p_i, \quad i\not= 1$
are even integers. The standard FQHE hierarchical theory
suggests  that at this filling the  FQHE state
is obtained by   condensation of quasiparticles of the previous levels.

The fractional fillings  observed to date occur
in the following sequence (filling $\nu <1$) \cite{jainc} :
\begin{eqnarray}
\nu & = & {n \over 2n+1}={1\over 3},{2\over 5}, {3\over 7},
\cdots , {9\over 19}, \cdots , \nonumber \\
\nu & = &1- {n \over 2n+1}={2\over 3},{3\over 5}, {4\over 7},
\cdots , {9\over 19}, \cdots ,  \nonumber \\
\nu & = & {n \over 4n+1}={1\over 5},{2\over 9}, {3\over 13},
\cdots ,  \nonumber \\
\nu & = &  1-{n \over 4n+1}={4\over 5}, {7\over 9},
\cdots , \nonumber \\
\nu & = &  {n \over 4n-1}={2\over 7},{3\over 11}, {4\over 15},
\cdots ,  \nonumber \\
\nu & = & 1- {n \over 4n-1}={5\over 7}, \cdots .
\label{sequ}
\end{eqnarray}
Consider in eq.\ (\ref{sequ}) the sequence of fillings
$\nu={(n / 2n+1)}, {n / (4n+1)}, {n / (4n-1)}$,
where the other fillings
correspond to the conjugate states. The filling
$\nu={n / (2n+1)}$ can be written in the form (\ref{shier})
with $p_1=3$ and $p_i=2, \quad i\not= 1$. Thus states at filling
$\nu={n / (2n+1)}$ are due to the hierarchical QE
condensations of the $\nu=1/3$
Laughlin state (in the way QE condense
mainly because of statistical interactions, as we discussed
previously).
FQHE states at such fillings are thus expected to be
more easily observed,
and it is actually so.

For $\nu={n / (4n+1)}$, $p_1=5$, and
$p_i =2, \quad i\not= 1$. Thus states at filling
${n / (4n+1)}$ are due to the hierarchical QE
condensations of the $\nu=1/5$ Laughlin state.
Since the $\nu=1/5$ filling of the parent state
is more difficult to observe
than the $\nu=1/3$ filling, so are states in the
sequence $\nu =n/(4n+1)$ compared with states in the
sequence $\nu =n/(2n+1)$.

$\nu={n / (4n-1)}$ can be written as
\begin{equation}
\nu ={1\over \displaystyle p_1+
{\strut 1\over \displaystyle p_2-
{\strut 1\over \displaystyle \cdots -
{\strut 1\over \displaystyle p_{n}}}}}   ,
\label{shier1}
\end{equation}
with $p_1=3$, $p_i=2, \quad i\not= 1$.
When $n=2$, the state is an hierarchical state due to the  QH
condensation   of the Laughlin state at filling
$1/3$ (the QH wavefunction  is
given by eq.\ (\ref{laugqh})).
When $n>2$, the states are due to the hierarchical QE condensation
of the hierarchical state at $n=2$.
It is found experimentally that
this sequence is  weaker than
the sequence $\nu = n/(2n+1)$, as it should from the preceding discussion
(sequences involving  QH condensations
should be weaker than sequences which do not).

It is interesting to observe  that QH condensation
only occurs in the second level of the hierarchical states
(or QH condensation only occurs
in parent Laughlin states at filling $1/p_1$).
The reason for this is possibly that in high levels
 of hierarchical states, QH
has a smaller electric charge and thus Coulomb interactions
are weaker. The hierarchical states
due to QH condensation in high levels
are rather difficult to observe.
However,  since the QE statistical interactions
play a significant role in producing the Laughlin states
 (\ref{laughqe}), it is still possible to
have hierarchical states due to QE condensations
in high level
hierarchical states despite their rather small charge.
In conclusion, the standard hierarchical theory
also explains the order of stability of sequences
found in experiments (as in Jain's theory).

Recent works show that the standard hierarchical theory
may be equivalent to Jain's theory \cite{he,yang}.
However, the construction of explicit trial
wavefunctions of electrons at hierarchical fillings
is still an open problem in the standard hierarchical theory.
In  Jain's theory, such trial wavefunctions
can be constructed without involving  quasiparticles.
Recent attempts have been made to get
the electron wavefunctions based on
the standard hierarchical theory \cite{martin},
but much more  remain to be done.

\vspace{1cm}

One of us (D.L) thanks for their hospitality
the theory division of IPN where part of this work was done.
He thanks Professors R. Ferrari and  S. Stringari
for their hospitality at Trento University where the work
was completed and Professor R. Iengo for useful discussions.
The work of D.L was partially supported by
ECC, Science Project SCI$^*$-CT92-0798.


\font\pet=cmr10 at 10truept
\font\bf=cmbx10 at 10truept

\pet


\begin{thebibliography}{999}
\baselineskip=12truept
\bibitem{lei}J.M. Leinaas and J. Myrheim, Nuovo Cimento
B{37} (1977) 1 ;
G.A. Goldin, R. Menikoff and D.H. Sharp, J. Math.
Phys. {22} (1981) 1664 ;
F. Wilczek, Phys.  Rev.  Lett.  {48} (1982) 1144;
{49}  (1982) 957;
Y.S. Wu, Phys.  Rev.  Lett. {52} (1984) 2103;
{53} (1984) 111.
\bibitem{halperin}
B.I. Halperin, Phys. Rev. Lett. {52} (1984) 1583.
\bibitem{haldane1}
F.D.M. Haldane, Phys. Rev. Lett. {51} (1983) 605.
\bibitem{pg}R. Prange and S. Girvin,  The Quantum Hall Effect
(Springer-Verlag, New York, Heidelberg,  1990, 2nd ed);
and references therein.
\bibitem{macdonald}A.H. MacDonald,
Perspective on the Quantum Hall Effect (Klewer, Boston, 1989);
M. Stone, Quantum Hall Effect (World Scientific, 1992);
and references therein.
\bibitem{arovas}D. Arovas, J.R. Schrieffer and F. Wilczek,
Phys. Rev. Lett. {53} (1984) 722.
\bibitem{wilczek}F. Wilczek,
Fractional Statistics and Anyon Superconductivity
(World Scientific, 1990); and references therein.
\bibitem{haldane}F.D.M. Haldane, Phys. Rev. Lett.
{67}  (1991) 937.
\bibitem{he}
S. He, X.C. Xie, and F.C. Zhang, Phys. Rev. Lett. {68} (1992) 3460;
M. Ma and F.C. Zhang, Phys. Rev. Lett. {66} (1991) 1769.
\bibitem{yang}
J. Yang and W.P. Su, Phys. Rev. Lett. {68} (1992) 2382;
Phys. Rev. Lett. {70} (1993) 1163;
Phys. Rev. B {47} (1993) 12953.
\bibitem{joh}M.D. Johnson and G.S. Canright, Florida
preprint UCF-CM-93-105.
\bibitem{ouvry}
A. Dasni\`{e}res de Veigy and S. Ouvry, Phys. Rev. Lett. {72 }
(1994) 600.
\bibitem{wu}Y.S. Wu,  ``Statistical Distribution for Particles Obeying
Fractional Statistics", Utah preprint.
\bibitem{jain}
J.K. Jain, Phys. Rev. Lett. {63} (1989) 199;
Phys. Rev. B {41} (1990) 7653;
Adv. phys. {41} (1992) 105.
\bibitem{dev}
G. Dev and J.K. Jain,  Phys. Rev. Lett. {69} (1992) 2843;
\bibitem{li}D. Li, Nucl. Phys. B {396} (1993)  411 (FS);
see also R. Iengo and K. Lechner, Phys. Rep. C{213} (1992) 179.
\bibitem{many}
G. V. Dunne, A. Lerda and C. A. Trugenberger,
      Mod. Phys. Lett. A {6}  (1991) 2891;
      Int. Jour. Mod. Phys. B {5} (1991)  1675;
G.V. Dunne, A. Lerda , S. Sciuto and C.A. Trugenberger,
      Nucl. Phys. B {370} (1992) 601;
J. Grundberg, T.H. Hansson, A. Karlhede and E. Westerberg,
Stockholm preprint USITP-91-2;
A.P. Polychronakos, Phys. Lett. B {264} (1991) 362;
C. Chou, Phys. Lett. A {155} (1991) 245;
Phys. Rev. D {44} (1991) 2533;
   A. Govari,  Technion preprint Phy.-92.
\bibitem{comtet}A. Comtet, J. McCabe and S. Ouvry,
Phys. Rev. D. {45} (1992) 709.
\bibitem{lispin}D. Li, Phys. Lett. A {169} (1992) 82.
\bibitem{gailuron} A. Dasni\a`eres de Veigy and S. Ouvry,
Phys. Lett. B {307 } (1993) 91.
\bibitem{iengoli}R. Iengo and D. Li,
Nucl. Phys. B  {413} (1994) 735 (FS).
\bibitem{egh}T. Eguchi, P.B. Gilkey and A.J. Hanson,
Phys. Rep. {66} (1980) 213.
\bibitem{laughlin}R.B. Laughlin, Phys. Rev. Lett. {50}  (1983) 1395.
\bibitem{jainc}J.K. Jain, Comments Cond. Mat. Phys. {16} (1993) 307.
\bibitem{martin}M. Greiter, Princeton  preprint IASSNS-HEP-92/78.
\end{thebibliography}
\end{document}